\documentclass{article}[11pt]
\usepackage{a4wide}
\usepackage{bm}
\usepackage{bbm}
\usepackage{epsfig,graphics}
\usepackage{amsmath,amssymb,amsfonts}
\usepackage{color}
\usepackage{epstopdf}
\usepackage[%dvipdfm,
colorlinks=true,
linkcolor=black,
breaklinks=true,
urlcolor=blue,
citecolor=green]{hyperref}
\usepackage{cite}

\newcommand{\order}[1]{\mathcal{O}\left(#1\right)}

\newcommand{\al}{&\!\!\!}

\newcommand{\decaytoK}{\Lambda_b^0\to K^- J/\psi p}
\newcommand{\decaytopi}{\Lambda_b^0\to \pi^- J/\psi p}

\begin{document}

\title{ Remarks on the $P_c$ structures and triangle singularities }

\author{Feng-Kun Guo$^{1,}$\footnote{Email address:
      \texttt{fkguo@itp.ac.cn} },~
      Ulf-G. Mei\ss ner$^{2,3,}$\footnote{Email address:
      \texttt{meissner@hiskp.uni-bonn.de} },~
      Juan Nieves$^{4}$\footnote{Email
      address: \texttt{jmnieves@ific.uv.es}}  ~and
      Zhi Yang$^{2,}$\footnote{Email address:
      \texttt{zhiyang@hiskp.uni-bonn.de} } \\
      {\it\small$^1$Key Laboratory of Theoretical Physics, Institute of
      Theoretical Physics, }\\
      {\it\small Chinese Academy of Science, Beijing 100190, China }\\
      {\it\small$^2$Helmholtz-Institut f\"ur Strahlen- und Kernphysik and Bethe
      Center for Theoretical Physics,}\\
      {\it\small Universit\"at Bonn, D-53115 Bonn, Germany}\\
      {\it\small$^3$Institute for Advanced Simulation, Institut f\"{u}r
       Kernphysik and J\"ulich Center for Hadron Physics,}\\
      {\it\small Forschungszentrum J\"{u}lich, D-52425 J\"{u}lich, Germany}\\
      {\it\small$^4$ Instituto de F\'isica Corpuscular (IFIC),
             Centro Mixto CSIC-Universidad de Valencia,}\\
       {\it\small    Institutos de Investigaci\'on de Paterna,
             Aptd. 22085, E-46071 Valencia, Spain}
      %\\
     % {\it\small$^4$ INPAC, Shanghai Key Laboratory for Particle Physics and
     % Cosmology, }\\
     % {\it\small Department of Physics and Astronomy, Shanghai Jiao-Tong
      % University, Shanghai 200240,   China}\\
       }
\date{\today}

\maketitle

\begin{abstract}

It was proposed that the narrow $P_c(4450)$ structure observed by the LHCb
Collaboration in the reaction $\Lambda_b\to J/\psi p K$  might be due to a
triangle singularity around the $\chi_{c1}$--proton threshold at 4.45~GeV. We
discuss the occurrence of a similar triangle singularity  in the $J/\psi
p$ invariant mass distribution for the decay $\Lambda_b\to J/\psi p \pi$, which
could explain the bump around 4.45~GeV in the data. More precise measurements of
this process would provide valuable information towards an understanding of the $P_c$
structures.

\end{abstract}

\medskip

\newpage

In 2015, the LHCb Collaboration reported the observation of two resonant-like
structures in the $J/\psi p$ invariant mass distribution in the decay process
$\Lambda_b\to K^- J/\psi p$~\cite{Aaij:2015tga}.
Fitting with Breit--Wigner forms, the masses and widths of these structures
are
\begin{eqnarray}
  M_{P_c(4380)} \al=\al (4380\pm8\pm29)~\text{MeV}, \qquad\qquad\!
  \Gamma_{P_c(4380)} = (205\pm18\pm86)~\text{MeV}, \nonumber\\
  M_{P_c(4450)} \al=\al (4449.8\pm1.7\pm2.5)~\text{MeV}, \qquad
  \Gamma_{P_c(4450)} = (39\pm5\pm19)~\text{MeV}\, .
\end{eqnarray}
Note that the signal for  the narrow $P_c(4550)$ is very clear, whereas the
necessity for including the broad $P_c(4380)$ may require some scrutiny (see
also the discussion in Ref.~\cite{Roca:2016tdh}).
If these structures  are genuine resonance states, being in the mass
region with a pair of charm and anti-charm quarks, they would contain five
valence quarks.
Further, the LHCb Collaboration recently reported a refined model-independent analysis
in Ref.~\cite{Aaij:2016phn} which shows that the $J/\psi p$
invariant mass distribution cannot be described without introducing additional
contribution due to exotic hadrons, such as the $P_c$, or rescattering effects,
such as the triangle singularities to be discussed here.
After the discovery, they have been suggested to be meson-baryon hadronic
molecules (predictions of such meson-baryon
states~\cite{Wu:2010jy,Wu:2010vk,Yang:2011wz,Xiao:2013yca} have been made a few
years before the LHCb discovery) or compact pentaquarks by a number of
groups~\cite{Chen:2015loa,Chen:2015moa,Roca:2015dva,Mironov:2015ica,
Maiani:2015vwa,He:2015cea,
Lebed:2015tna,Meissner:2015mza,Li:2015gta,Wang:2015epa,Scoccola:2015nia,
Anisovich:2015zqa,Wang:2015qlf,Yang:2015bmv,Lu:2016nnt,Shimizu:2016rrd,
Shen:2016tzq,Santopinto:2016pkp}
(see Refs.~\cite{Burns:2015dwa,Chen:2016qju} for partial reviews).

However, not all peaks in invariant mass distributions are due to resonances,
which are poles of the $S$-matrix. The $S$-matrix also possesses other types of
singularities, such as the branch points (and the associated cuts) at two-body
thresholds, and also the so-called triangle singularity originating from a triangle diagram.
Poles of the $S$-matrix are of dynamical origin in the sense that they exist
because the interaction among the internal constituents is strong enough
such that poles are generated in the scattering amplitude. This is necessarily a non-perturbative
phenomenon. In contrast, singularities like branch points and triangle singularities are of
kinematical origin. They emerge in the physical amplitude and can produce observable effects when
the kinematics of a process is special. For detailed discussions about the triangle
singularity, we refer to the 1960ties monographs~\cite{Eden:book,Chang:book},
recent lecture notes by one of the key players in the old days~\cite{Aitchison:2015jxa}
and the book~\cite{Anisovich:2013gha}.
For instance, considering a triangle diagram, if the involved masses and momenta
are such that all of the three intermediate particles can go on shell with all
of the interaction vertices satisfying energy-momentum
conservation~\cite{Coleman:1965xm}, then the triangle singularity is on the
physical boundary and can show up as a peak in the corresponding  invariant mass
distribution. %
%%%%%%%%%%%%%%%%%%%%%%%%%%%%%%%%%%%%%%%%%%%%%%%%%%%%%%%%%%%%%%%%%%%%%%%%%%%%%%%%
\begin{figure}[tbh]
  \centering
    \includegraphics[width=0.4\linewidth]{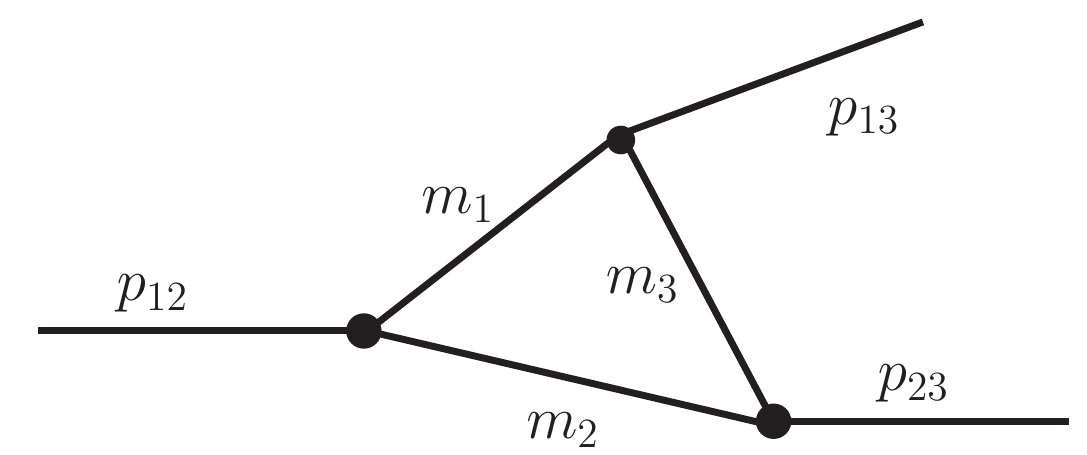}
  \caption{A triangle diagram. Here,  each internal line is labelled by the mass of
  the corresponding particle and each external line by its four-momentum.
  }
  \label{fig:triangle}
\end{figure}
%%%%%%%%%%%%%%%%%%%%%%%%%%%%%%%%%%%%%%%%%%%%%%%%%%%%%%%%%%%%%%%%%%%%%%%%%%%%%%%%
To be more explicit, let us take Fig.~\ref{fig:triangle} and explain the
kinematical region where the triangle singularity can occur. The diagram can be
interpreted as particle $p_{12}$ decays into particles $m_1$ and $m_2$,
following by the sequential decay of $m_1$ into $p_{13}$ and $m_3$, and $m_2$
and $m_3$ react to generate the external $p_{23}$. We may consider the rest frame of
$p_{12}$. The triangle singularity is on the physical boundary when all of the
intermediates states $m_{1,2,3}$ are on their mass shell, and moreover,
particle $m_3$ moves along the same direction and with a larger speed than
particle $m_2$. In this kinematical region, particle $m_3$ can catch up with
particle $m_2$ to rescatter like a classical process.

  In the last few years,
triangle singularities have been used to explain various structures in invariant
mass distributions of two or three
hadrons~\cite{Wu:2011yx,Wu:2012pg,Wang:2013hga,Ketzer:2015tqa,
Achasov:2015uua,Lorenz:2015pba,Szczepaniak:2015eza}. Triangle singularities
present another possibility of explaining the LHCb $P_c$ structures.
In particular,  the narrower structure $P_c(4450)$ coincides with
the $\chi_{c1}\,p$ threshold which is also the position of the triangle
singularity of the loop diagram with $\Lambda^*(1890)\,\chi_{c1}\,p$
intermediate states~\cite{Guo:2015umn,Liu:2015fea}.
% {\color{red}{No matter what the $P_c$ structures are, the triangle
% singularities exists there. But their contribution to the observed peaks is
% unknown due to the lack of knowledge on the interaction strengths in the loop.
% If the $P_c$ structures are really resonances, the study of triangle
% singularities is also important to extract their masses and widths from the
% experimental data.}}
If the $P_c$ structures are really due to triangle singularities, the LHCb
discovery will still be very important because triangle singularities are one of
the intriguing analytic properties of the $S$-matrix, and their observability
has been discussed since long time ago (see Ref.~\cite{Schmid:1967} for a
classics, and Ref.~\cite{Liu:2015taa} for a recent discussion).

Being the first candidates of explicitly exotic pentaquark states\footnote{In
fact, the quantum numbers of the $P_c$ structures can be formed by three light
quarks, and thus  are non-exotic.
However, since they are located above 4~GeV, if they are light baryons the vast
amount of phase space would allow them to decay into light hadrons very fast and
the widths would be much larger than those reported by the LHCb Collaboration.
Therefore, it is more natural to assume that there are a pair of charm and
anticharm quarks inside the annihilation of which into light hadrons are
suppressed.}, either as hadronic molecules or compact pentaquarks, one of the
utmost important issues regarding the observed $P_c$ structures is to
distinguish the kinematic singularity explanation from resonances.
Because the triangle singularity depends on the kinematics very strongly, one
should search for the $P_c$ structures in processes with a different kinematics
where the triangle singularities discussed in
Refs.~\cite{Guo:2015umn,Liu:2015fea} do not play a role. For this purpose,
suggested experiments include searching for the $P_c$ structures in the
$\chi_{c1} p$ invariant mass distribution of the decay $\Lambda_b\to
K^-\chi_{c1} p$ which is expected to have a similar branching fraction as that
of the $\Lambda_b\to K^-J/\psi p$~\cite{Guo:2015umn}, and searching for the
structures in reactions with a different kinematics such as the photoproduction
processes~\cite{Wang:2015jsa,Kubarovsky:2015aaa,Karliner:2015voa,Huang:2016tcr},
pion induced reactions~\cite{Lu:2015fva,Liu:2016dli} and heavy ion
collisions~\cite{Wang:2016vxa,Schmidt:2016cmd}.
From this point of view, investigating whether other processes with  a $J/\psi
p$ in the final state have similar triangle singularities is important and
necessary. One of such processes which is closely related to the observation
process of the $P_c$ structures is the reaction $\Lambda_b\to \pi^- J/\psi p$,
the $J/\psi p$ invariant mass distribution of which has in fact been measured by
the LHCb Collaboration~\cite{Aaij:2014zoa}. This process was studied in
Refs.~\cite{Wang:2015pcn} in the context of hidden-charm pentaquarks, and it is
noted that there is indeed a nontrivial structure around 4.45~GeV.
% , which was used to challenge the triangle singularity explanation of
% Ref.~\cite{Guo:2015umn}.

Here we will discuss further the model proposed in Ref.~\cite{Guo:2015umn} which
observes that the narrow $P_c(4450)$ might be due to kinematic singularities
related to the $\chi_{c1}p$ normal and abnormal thresholds. In
Ref.~\cite{Guo:2015umn}, it is noticed that the narrow $P_c(4450)$ is located
exactly at the $\chi_{c1}p$ threshold, $M_{P_c(4450)} - M_{\chi_{c1}} - M_p =
(0.9\pm3.1)~\text{MeV}$, and when the measured decay $\Lambda_b\to K^- J/\psi p$
occurs through the $\Lambda^*(1890) \chi_{c1}p$ triangle diagram, where the
$\Lambda_b$ decays into the $\Lambda^*(1890)$ and $\chi_{c1}$ first and the
proton as a decay product of the $\Lambda^*(1890)$ rescatters  with the
$\chi_{c1}$ into the $J/\psi p$, the triangle singularity (abnormal threshold)
is also located at the same place. To be more precise, it is slightly moved into
the complex energy plane because the $\Lambda^*(1890)$ can decay into $K^-p$ and
thus has a finite width. We will show that there can also be triangle
singularities around 4.45~GeV in the three-body decay $\Lambda_b\to \pi^- J/\psi
p$ as well, which are able to reproduce the nontrivial $J/\psi p$ invariant mass
distribution in that region.

For a given triangle diagram, the locations of the triangle singularities, also
called leading Landau singularities, are given by solving the quadratic
equation~\cite{Landau:1959fi}
\begin{equation}
    1 + 2\, y_{12}\, y_{23}\, y_{13} = y_{12}^2 + y_{23}^2 + y_{13}^2,
    \label{eq:Landau}
\end{equation}
with $y_{ij} = (m_i^2 + m_j^2 - p_{ij}^2)/(2\,m_i\,m_j)$. The definitions of the
momenta and masses can be read off from Fig.~\ref{fig:triangle}. The
singularities need to be understood with the help of the analytic properties of the
$S$-matrix. For instance, let us consider the diagram shown in
Fig.~\ref{fig:triangle}, and consider the case measuring the $s_{23}\equiv
p_{23}^2$ invariant mass distribution. The threshold of $m_2$ and $m_3$ leads to a
branch point. The cut from this branch point divides the complex $s_{23}$
plane into two Riemann sheets, and the upper boundary of the cut in the first
Riemann sheet presents the physical region of the process $m_2+m_3\to p_{23}$,
where $p_{23}$ can be regarded as the total momentum of the particles emitted
from the scattering between $m_2$ and $m_3$.  It turns out that one of the
solutions of the quadratic equation in Eq.~\eqref{eq:Landau} is always far from
the physical region, and the the other one could be close to the physical region
(and also  close to the threshold of $m_2$ and $m_3$).
This happens when it is located in the lower half plane of the second Riemann sheet, below the cut,
with a small imaginary part (for a more detailed discussion for the case of the
$P_c (4550)$, see Ref.~\cite{Guo:2015umn}). The condition is that all of the three
intermediate particles are on their mass shells, and the rescattering between
$m_2$ and $m_3$ can happen as a classical process~\cite{Coleman:1965xm}. It is
fulfilled when $m_1$ is inside or at least in the vicinity of the range $m_1 \in \left[
m_{1,\text{low}}, m_{1,\text{high}} \right]$ with~\cite{Guo:2015umn,Liu:2015taa}
\begin{equation}
   m_{1,\text{low}} =\sqrt{\frac{p_{12}^2 m_3 + p_{13}^2 m_2}{m_2
 + m_3}  - m_2 m_3 }
 \quad\text{and}\quad
 m_{1,\text{high}} = \sqrt{p_{12}^2} - m_2 \,.
 \label{eq:range}
\end{equation}
Substituting this range of $m_1$ into Eq.~(\ref{eq:Landau}), and noticing that
only one of the solutions is possible to be near the physical region~(see
discussions in, e.g., Refs.~\cite{Aitchison:1964zz,Schmid:1967,Guo:2015umn}),
the triangle singularity in the variable $p_{23}^2$ lies in the corresponding
range
\begin{equation}
  p_{23,\text{\,sing.}}^2 \in \left[ (m^{}_2+m^{}_3)^2,
  %\sqrt{p_{12}^2} m_2 +
  %\frac{\sqrt{p_{12}^2} m_3^2- m_2 p_{13}^2}{\sqrt{p_{12}^2}-m_2}
  m_2^2 + m_3^2 - 2 m^{}_2 m^{}_3 y^{}_{23}
  \right] .
\end{equation}
The upper bound can also be written as $m_2^2 + m_3^2 + 2 m^{}_2 m^{}_3
y^{}_{13}$ by using that $y_{12}=-1$ and $y_{23}+y_{13}=0$, which are valid only
for $m_1=m_{1,\text{high}}$.
One notices that if the resonance $m_1$ takes a mass of $m_{1,\text{low}}$, the
amplitude will be singular at $p_{23}^2=(m_2+m_3)^2$. Of
course, a physical amplitude never diverges in the physical region. In this
case, the fact that all the intermediate particles can go on shell means that
the particle $m_1$ can decay into particles $m_3$ and $p_{13}$, and thus it must
be an unstable resonance. As a consequence, the triangle singularity cannot
reside on the real $p_{23}^2$ axis, but in the complex plane so that
the relevant amplitude in the physical region still takes a finite value.
Nevertheless, if the singularity is not located deep in the complex plane, it will then
introduce a visible peak around the real part of the singularity location in the
$p_{23}^2$ distribution. In addition, for a value of $m_1$ slightly beyond the
range given in Eq.~\eqref{eq:range}, the singularity is not far from the
physical boundary and could still cause a visible effect.

\begin{figure}[tbh]
  \centering
    \includegraphics[width=0.8\linewidth]{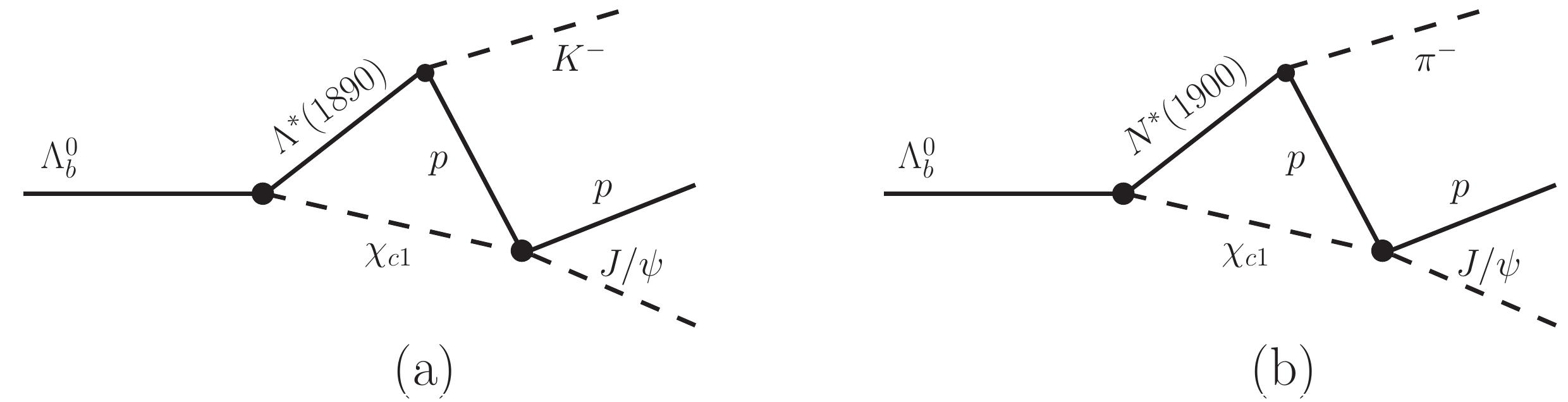}
  \caption{Triangle diagrams which can produce a peak around 4.45~GeV in the
  $J/\psi p$ invariant mass distribution for the processes (a) $\decaytoK$ and
  (b) $\decaytopi$.
  }
  \label{fig:loops}
\end{figure}

Since the peak position of the $P_c(4450)$ coincides with the $\chi_{c1}p$
threshold, we consider the case with $m_2$ and $m_3$ being the $\chi_{c1}$ and
proton, respectively. The triangle diagram suggested in Ref.~\cite{Guo:2015umn}
for the process $\decaytoK$ is shown in Fig.~\ref{fig:loops}~(a), with the mass
of the $\Lambda^*(1890)$ exactly at the lower edge of the range
$[1.89,2.11]$~GeV given by Eq.~\eqref{eq:range}.
% %%%%%%%%%%%%%%%%%%%%%%%%%%%%%%%%%%%%%%%%%%%%%%%%%%%%%%%%%%%%%%%%%%%%%%%%%%%%%%%
% %%%%%%%%%%%%%%%%%%%%%%%%%%%%%%%%%%%%%%%%%%%%%%%%%%%%%%%%%%%%%%%%%%%%%%%%%%%%%%%
Replacing the $\Lambda^*$ by an $N^*$ resonance, we get the analogue for the
process $\decaytopi$. The relevant mass range of the $N^*$ is $[1.84,2.11]$~GeV.
Within this range, there are two three-star nucleon resonances:
the $N^*(1875)$ with $J^P=\frac32^-$ and the $N^*(1900)$ with $J^P=\frac32^+$.
Substituting $(1.875-i\,0.125)$~GeV and $(1.9-i\,0.1)$~GeV as their
masses~\footnote{Here, the values refer to $M-i\,\Gamma/2$, and we use the
central values of the masses and widths as given by the
PDG~\cite{Agashe:2014kda}.} in Eq.~\eqref{eq:Landau}, we find triangle
singularities at
\begin{equation}
(4429-i\,10)~\text{MeV\quad and \quad}~(4439-i\,16)~\text{MeV}~ ,
\label{eq:singvalue}
\end{equation}
respectively.
Because the singularity is in
the second Riemann sheet of the complex $m_{J/\psi p}$ plane, the absolute value
of amplitude with the singularities as given in Eq.~\eqref{eq:singvalue}, as well as
that for the $\decaytoK$, is maximized at the $\chi_{c1}p$
threshold. This is because the real parts of the singularity
positions are smaller than the branch point, the $\chi_{c1}p$ threshold. It
is thus similar to the case that an amplitude that possesses a virtual state pole
has a sharp cusp at the relevant threshold. However, since the imaginary parts
of the values given above is larger than that for the one induced by the
$\Lambda^*(1890)$ for the process $\decaytoK$, $(4447.8-i\,0.3)$~MeV, the peak
due to the triangle singularities through the exchange of the $N^*$ as shown in
Fig.~\ref{fig:loops}~(b) should have a larger width than that for
Fig.~\ref{fig:loops}~(a).
%This is in fact what was observed in the data of the
%$J/\psi p$ invariant mass distribution for the $\decaytopi$ process.

\begin{figure}[tbh]
  \centering
    \includegraphics[width=0.49\linewidth]{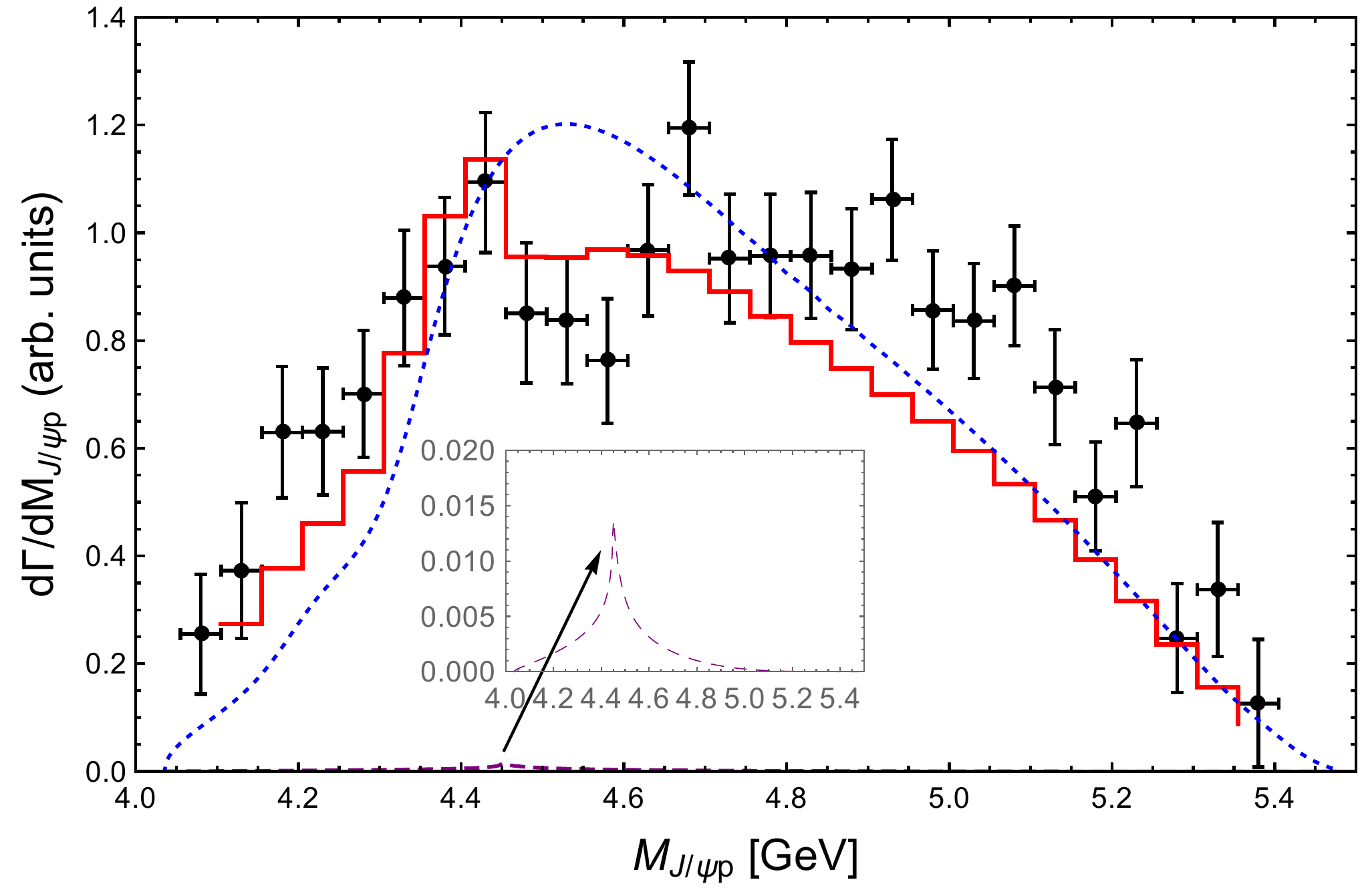}\hfill
    \includegraphics[width=0.49\linewidth]{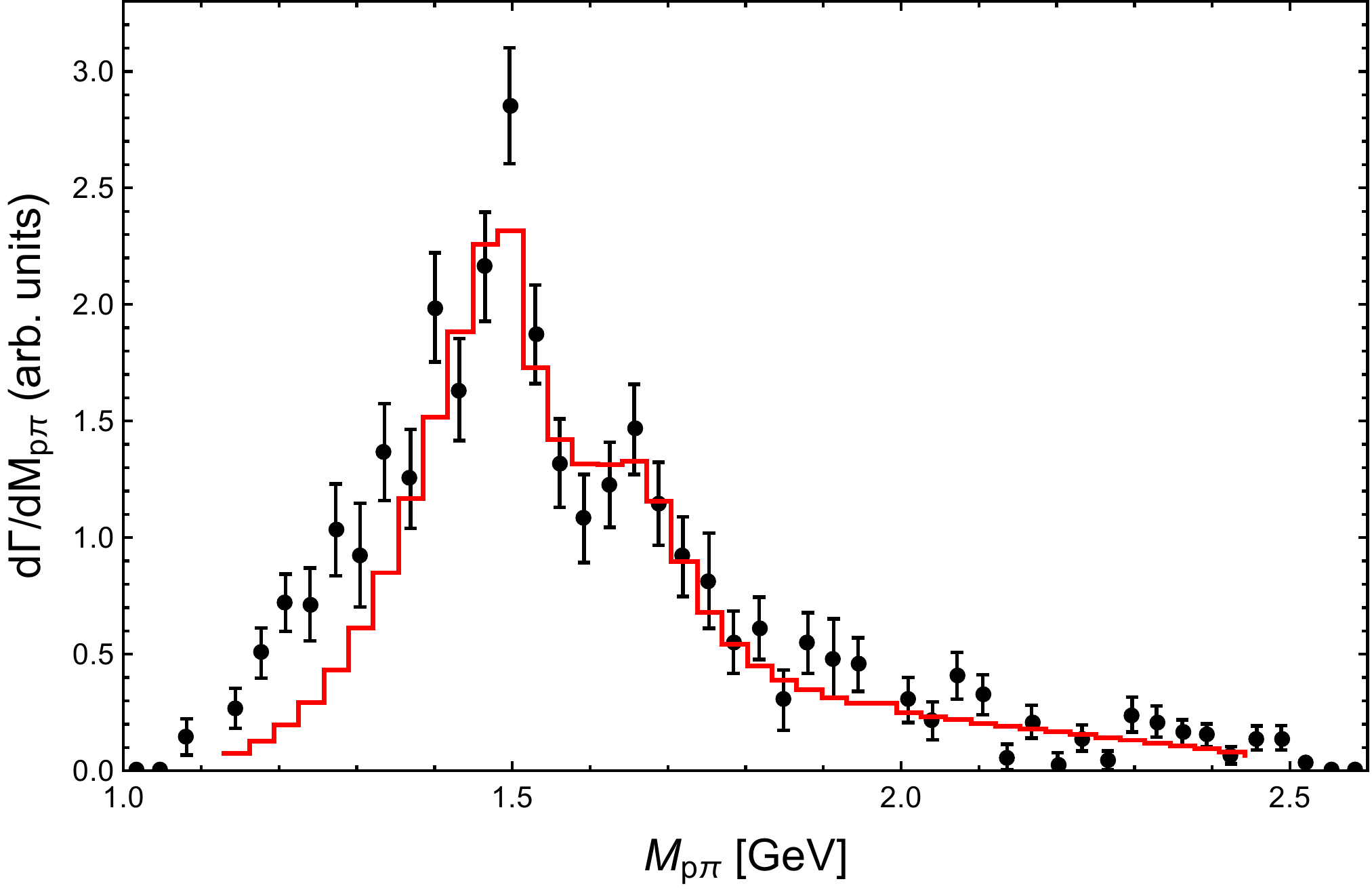}
  \caption{Comparison of a model including the triangle singularities from
  Fig.~\ref{fig:loops}~(b) with the experimental data. Tree-level exchanges of the
  $N^*(1440)$, the $N^*(1520)$ and the $N^*(1650)$, as depicted in
  Fig.~\ref{fig:diagrams}, are included to describe the $p\pi$ invariant mass
  distribution. The data are taken from Ref.~\cite{Aaij:2014zoa}. The solid
  lines and binned histograms show the fit to the data in the range
  $M_{J/\psi p}\in [4.33,4.58]$~GeV and $M_{p\pi}\in [1.33,1.85]$~GeV,
  the dotted line shows the
  contribution from the tree-level exchange of the $N^*$ resonances and
  the dashed line show the contribution from triangle diagram.}
  \label{fig:comparison}
\end{figure}

Here, we do not intend to construct a full model for the three-body decay
$\decaytopi$, which is a formidable task if all the final
state interactions including the exchange of $N^*$ resonances, and even the
exotic $Z_c(3900)$, and kinematical singularities are taken into account.
Instead, we only want to illustrate that the bump around 4.45~GeV in its $J/\psi
p$ invariant mass distribution observed by the LHCb collaboration may be due to the triangle
singularities discussed above. Since we do not know the relative strength for the decays
$\Lambda_b\to N^*(1875)\chi_{c1}$ and $\Lambda_b\to
N^*(1900)\chi_{c1}$, we choose to include only the $N^*(1900)$ which has the
same spin and parity as the $\Lambda^*(1890)$. We also include as an additional
contribution the tree-level exchanges of the $N^*(1440)$, the $N^*(1520)$ and the
$N^*(1650)$ with the masses and widths taken from Ref.~\cite{Agashe:2014kda}.
The $N^*$ exchanges can describe well the $p\pi$ invariant mass distribution, and provide a smooth
background to the $J/\psi p$ one.  It is its interference with the
triangle singularity that produces the observed peak around 4.4~GeV in our fit
as shown in Fig.~\ref{fig:comparison}.

%%%%%%%%%%%%%%%%%%%%%%%%%%%%%%%%%%%%%%%%%%%%%%%%%%%%%%%%%%%%%%%%%%%%%%%%%%%%%%%%
%%%%%%%%%%%%%%%%%%%%%%%%%%%%%%%%%%%%%%%%%%%%%%%%%%%%%%%%%%%%%%%%%%%%%%%%%%%%%%%%
%%%%%%%%%%%%%%%%%%%%%%%%%%%%%%%%%%%%%%%%%%%%%%%%%%%%%%%%%%%%%%%%%%%%%%%%%%%%%%%%
\begin{figure}[tbh]
  \centering
    \includegraphics[width=0.3\linewidth]{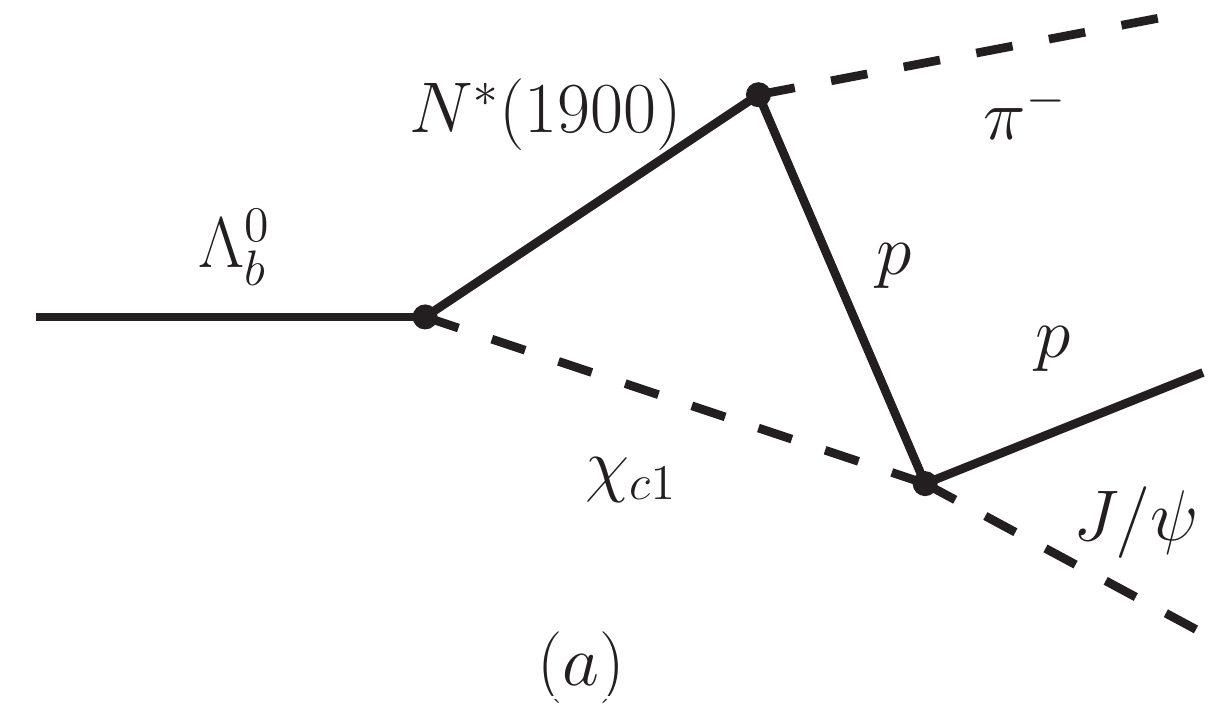}\hfill
    \includegraphics[width=0.85\linewidth]{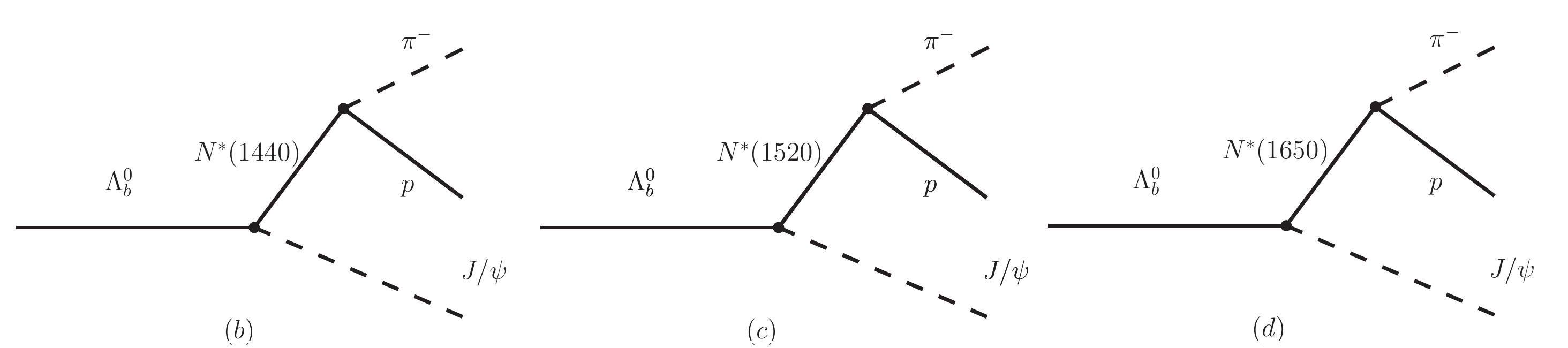}
  \caption{Diagrams in the simple model used to fit to the LHCb data.}
  \label{fig:diagrams}
\end{figure}
%%%%%%%%%%%%%%%%%%%%%%%%%%%%%%%%%%%%%%%%%%%%%%%%%%%%%%%%%%%%%%%%%%%%%%%%%%%%%%%%

%%%%%%%%%%%%%%%%%%%%%%%%%%%%%%%%%%%%%%%%%%%%%%%%%%%%%%%%%%%%%%%%%%%%%%%%%%%%%%%%
\begin{figure}[tbh]
  \centering
    \includegraphics[width=0.6\linewidth]{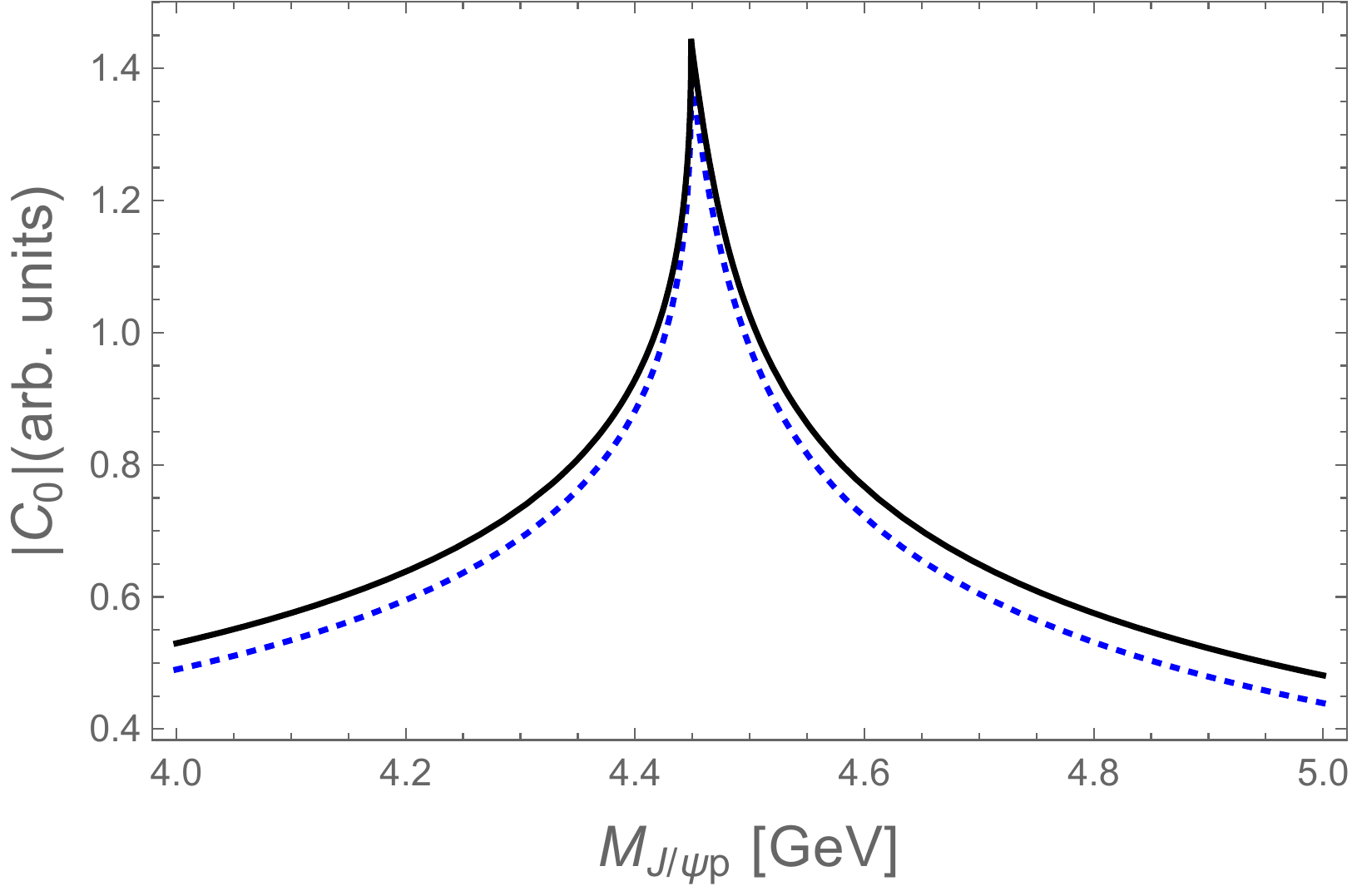}\hfill
  \caption{Comparison of the scalar three-point loop function for the
  $N^*(1900)$ with a complex mass $(1.9-i\,0.1)$~GeV (solid curve) with that
  convoluted with the spectral function as given in Eq.~\eqref{eq:spectral}
  (dashed curve).}
  \label{fig:spectral}
\end{figure}
%%%%%%%%%%%%%%%%%%%%%%%%%%%%%%%%%%%%%%%%%%%%%%%%%%%%%%%%%%%%%%%%%%%%%%%%%%%%%%%%
In the fit, we have used a complex mass, $M-i\,\Gamma/2$, to take
the width effect of $N^*(1900)$ into account. Although this is not the rigorous
way to deal with the width effect, such a method has been shown to be able
to correctly account for the peak effects in Ref.~\cite{Aitchison:1964zz}. In order to check
that statement, we compare the absolute value of the scalar triangle loop
function, denoted as $C_0(m_{N^*}^2)$, for the $N^*(1900)$ with a complex mass
$m_{N^*}=(1.9-i\,0.1)$~GeV with that with a variable mass convoluted with a
spectral function as follows:
\begin{equation}
  \frac1{\pi}\int_{s_a}^{s_b} ds\, \text{Im}\left(
  \frac{-1}{s-m^2+i\,m\,\Gamma}\right) C_0(s)\, ,
  \label{eq:spectral}
\end{equation}
with $m=1.9$~GeV and $\Gamma=0.2$~GeV where the integration region is taken to
be between $s_a=(m-2\,\Gamma)^2$ and $s_b=(m+2\,\Gamma)^2$.
The comparison is shown in Fig.~\ref{fig:spectral} by the solid and dashed
curves showing the results with a complex mass and with a spectral function as
given above, respectively.
One sees that indeed the peaks in both cases are quite similar to each other.

%   \ldots The second peak
% in the $J/\psi p$ invariant mass distribution is produced by a triangle
% singularity as well, and the responsible diagram is given in
% Fig.~\ref{fig:diagrams} (to be provided by Zhi).
%

\begin{figure}[tbh]
  \centering
    \includegraphics[width=0.7\linewidth]{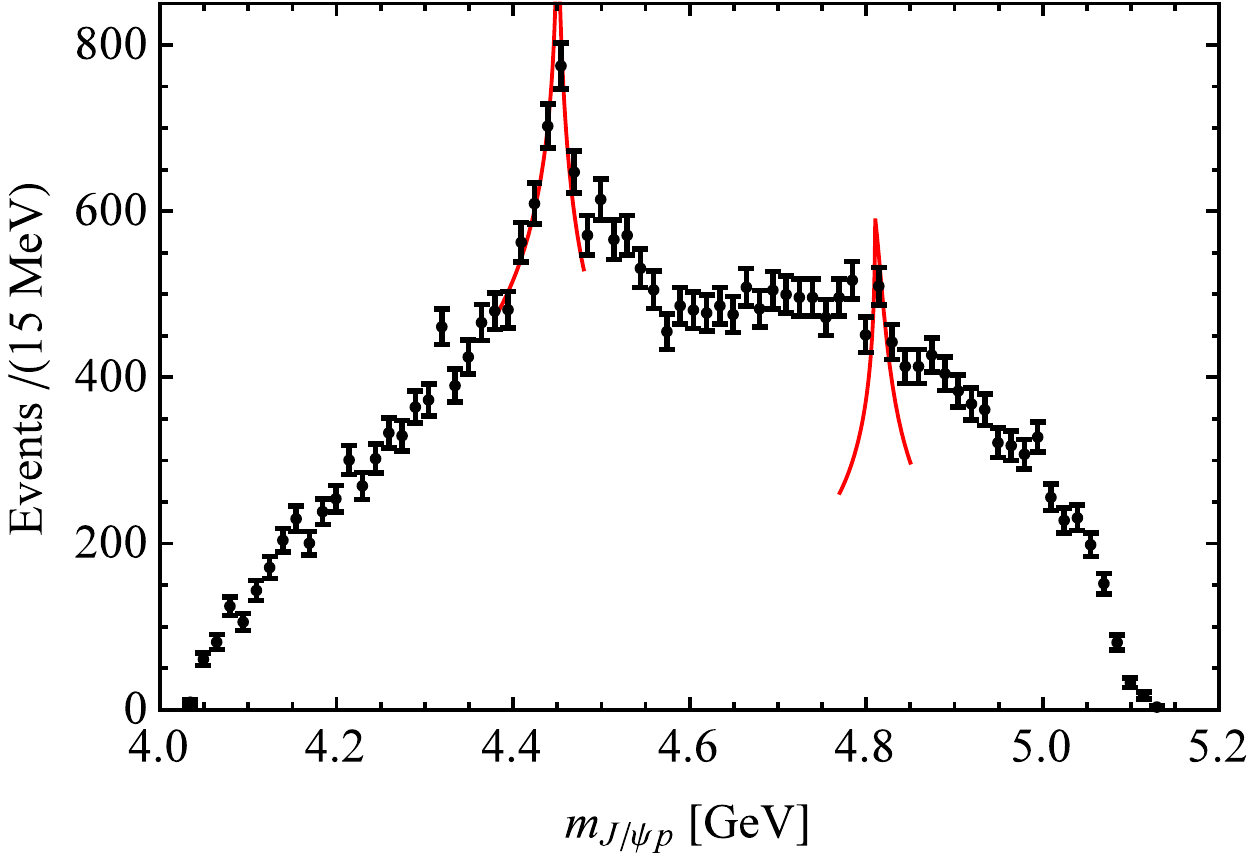}
  \caption{Comparing the locations of two triangle singularities close to the
  $\chi_{c1}p$ and $X(3872)p$ thresholds, respectively, with the LHCb data of
  the $J/\psi p$ invariant mass distribution for the decay $\decaytoK$.  }
  \label{fig:comparison1}
\end{figure}

Because the $\decaytopi$ decay is Cabibbo-suppressed in comparison with the
$\decaytoK$ decay, the observed number of events is smaller, and as a result,
the bin width of the reported data of the $J/\psi p$ invariant mass distribution
is 50~MeV in Ref.~\cite{Aaij:2014zoa}, while it is  15~MeV in
Ref.~\cite{Aaij:2015tga} where the $P_c$ structures were observed.
 If the $P_c(4450)$ is really due to triangle singularities, one
would expect that it behaves differently in different reactions as different
intermediate states are involved. One sees that the dashed line in the left
panel of Fig.~\ref{fig:comparison} is broader than the peak around 4.45~GeV in
Fig.~\ref{fig:comparison1}. On the contrary, if the $P_c(4450)$ is due to a real
resonance, one would expect it to have the same width in the same final states
$J/\psi\pi$ for both decays of $\decaytopi$ and $\decaytoK$. However, with a
50~MeV bin width, one cannot distinguish between these two scenarios. It is
thus important to measure the $\decaytopi$ more precisely to distinguish the
triangle singularity from the resonance scenario.

We want to emphasize here that the mechanism in this model should not be
regarded as a complete model for the decay $\decaytopi$. Its sole purpose is to
show that the peaks as observed are compatible with those produced by triangle
singularities. Therefore,  the confirmation of the $P_c$ structures in
other processes with very different kinematics which are expected to be free of
the singularities discussed here, is urgently called for.

There could be even  more triangle singularities. For instance, replacing the
$\chi_{c1}$ by the $X(3872)$ and replacing the $\Lambda^*(1890)$ by a
$\Lambda^*$ with a mass in the range from 1.65~GeV to 1.75~GeV, like e.g. the four-star
$J^P=3/2^-$ resonance $\Lambda^*(1690)$, one would be able to get a triangle
singularity close to the threshold of the $X(3872)$ and proton at about
4.81~GeV. A comparison of the locations of the triangle singularity discussed in
Ref.~\cite{Guo:2015umn} and that due to the singularity of the
$\Lambda^*(1690)$--$X(3872)$--$p$ triangle diagram with the LHCb data of the
$J/\psi p$ invariant mass distribution for the decay $\decaytoK$ is presented
in Fig.~\ref{fig:comparison1}, where the solid lines with sharp peaks are given
by the absolute value of the corresponding triangle diagram. We remark
that the purpose here is only to compare the locations of the triangle
singularities in the $J/\psi p$ invariant mass distribution with the data.

%%%%%%%%%%%%%%%%%%%%%%%%%%%%%%%%%%%%%%%%%%%%%%%%%%%%%%%%%%%%%%%%%%%%%%%%%%%%%%%%
%%%%%%%%%%%%%%%%%%%%%%%%%%%%%%%%%%%%%%%%%%%%%%%%%%%%%%%%%%%%%%%%%%%%%%%%%%%%%%%%

For a quantitative understanding of the contribution of the triangle
singularities discussed here and in Ref.~\cite{Guo:2015umn}, the couplings for
all the three vertices need to be known. While the pionic (kaonic) decays from
the $N^*$ ($\Lambda^*$) to the nucleon can be obtained from the corresponding
partial decay widths of the  $N^*$ ($\Lambda^*$), the weak decays of the
$\Lambda_b$ into the $\chi_{c1}$ and $N^*$ ($\Lambda^*$) need to be measured. In
addition, lattice QCD calculations are also necessary for obtaining reliable
information on  the rescattering $\chi_{c1}p\to J/\psi p$.
Because there is no common valence quark between a charmonium and a nucleon, the
rescattering process $\chi_{c1} p \to J/\psi p$ is not expected to be very
strong.\footnote{It is worthwhile to notice that such a meson--baryon
interaction scales as $\order{1/N_c}$ in the large $N_c$
limit~\cite{Witten:1979kh}, not as much suppressed as the analogous
OZI--forbidden meson--meson scattering which scales as $\order{1/N_c^{2}}$. }
However, we notice that a recent lattice QCD
calculation by the NPLQCD Collaboration reveals possible existence of
charmonium-nucleus bound states~\cite{Beane:2014sda}. Note that the lattice
calculation was performed at a pion mass as heavy as 805~MeV. A leading order
chiral extrapolation to the physical quark masses results in a charmonium-nucleus
binding energy of $\lesssim40$~MeV~\cite{Beane:2014sda}.

To summarize, we have shown that there could be a triangle singularity around
the $\chi_{c1}p$ threshold in the measured $J/\psi p$ invariant mass
distribution for the decay $\decaytopi$. It originates from a triangle diagram with
the $\chi_{c1}$, proton and a narrow $N^*$ resonance, in the mass range of
[1.84,\,2.11]~GeV, as intermediate states. It is similar to the model explaining
the $P_c(4450)$ in the decay $\decaytoK$ proposed in Ref.~\cite{Guo:2015umn},
where the $\Lambda^*(1890)$ takes the place of the $N^*$.
More precise
measurements of the $\decaytopi$ will be extremely helpful to study the $P_c$
structures.
% {\color{red}{
% The observed peak in the left diagram of Fig.~\ref{fig:comparison}
% is much broader than that in Fig.~\ref{fig:comparison1}, this is consistent with
% our expectation from the triangle singularity. However, there are large uncertainties on
% the current measurement of the process $\decaytopi$, thus more precise
% measurements will be extremely helpful to study the $P_c$ structures.}}

\medskip

{\bf Note added:}  After the submission of this manuscript to the journal,
LHCb reported on a new measurement on the $\decaytopi$~\cite{Aaij:2016ymb}.
The $P_c$ structures are not as significant as those in the
$\decaytoK$~\cite{Aaij:2015tga}. Yet, it was claimed that consistent production
rates in both processes were obtained.

\medskip

\section*{Acknowledgments}

We would like to thank Eulogio Oset for helpful discussions.
FKG gratefully acknowledges the hospitality at the HISKP where part of this work
was performed. UGM and ZY acknowledge the hospitality of the ITP/CAS, where part
of this work was done. This work is supported in part by DFG and NSFC through
funds provided to the Sino-German CRC 110 ``Symmetries and the Emergence of
Structure in QCD'' (NSFC Grant No.~11621131001),  by the Thousand Talents Plan
for Young Professionals, by the Chinese Academy of Sciences (CAS) (Grant No.
QYZDB-SSW-SYS013), by the CAS President's International Fellowship Initiative
(PIFI) (Grant No. 2015VMA076), by the Spanish Ministerio de Econom\'\i a y
 Competitividad and European FEDER funds under the contracts
 FIS2014-51948-C2-1-P,  FIS2014-57026-REDT and SEV-2014-0398  and by Generalitat
 Valenciana under contract PROMETEOII/2014/0068.

\end{document}